\documentclass[aps,prc,twocolumn,superscriptaddress,nofootinbib,longbibliography]{revtex4-2}

\usepackage{graphicx}
\usepackage{amsmath}
\usepackage{hyperref}
\usepackage{color}
\usepackage{xspace}

\newcommand{\Ntrk}{\ensuremath{N_{\mathrm{trk}}}\xspace}
\newcommand{\Ntrkcut}{\ensuremath{N_{\mathrm{trk}}^\mathrm{cut}}\xspace}

\newcommand{\Eta}{\Eta}

\begin{document}

\title{Data-driven method to estimate contamination from light ion beam transmutation at colliders}

\author{Sruthy Jyothi Das}
\email{sjdas@uic.edu}

\author{Austin Baty}
\email{abaty@uic.edu}
\affiliation{Department of Physics, University of Illinois Chicago, Chicago, Illinois, USA}

\date{\today}

\begin{abstract}
Collisions of relativistic light ions such as oxygen, neon, and magnesium, have been proposed as a way to examine the system-size dependence of dynamics typically associated with the quark-gluon plasma produced in collisions of heavier ions such as xenon, gold, or lead.  Recent efforts at both the Relativistic Heavy Ion Collider (RHIC) and Large Hadron Collider (LHC) have produced large datasets of proton-oxygen, oxygen-oxygen, and neon-neon collisions, catalyzing intense interest in experimental backgrounds associated with light ion collisions. In particular, electromagnetic dissociation of light ions while they are circulating in a collider can result in beam contamination that is difficult to simulate precisely.  Here we propose a data-driven method for evaluating the potential impact of beam contaminants on physics analyses.  The method exploits the time dependence and smaller size of contaminant ion species to define control regions that can be used to quantify potential contamination effects.  A simple model is used to illustrate the method and to study its robustness.  This method can inform studies of recent LHC and RHIC data and could also be useful for future light ion programs at the LHC and beyond.

\end{abstract}

\maketitle

\section{Introduction}

In the field of high energy nuclear physics, collisions of relativistic heavy ions, e.g., xenon, gold, or lead, are used to study the emergent properties of the strong force.  In these collisions a strongly interacting hot medium, known as quark-gluon plasma (QGP), is created~\cite{Busza:2018rrf,Niida:2021wut,Harris:2023tti}.  The QGP exhibits many interesting and nontrivial dynamics including collective fluid-like behavior~\cite{Romatschke:2017ejr,Heinz:2013th} and results in strong modifications to particle production as compared to proton-proton collision baselines~\cite{CMS:2016uxf,ATLAS:2014ipv,ALICE:2013dpt}. 

Perhaps surprisingly, critical examination of other collision systems having a much smaller size, such as proton-gold, deuteron-gold, helium-3-gold~\cite{PHENIX:2017xrm}, and proton-lead collisions~\cite{ATLAS:2016yzd,CMS:2015yux} have revealed similar behaviors as heavy ion collisions, prompting the question as to whether a QGP droplet is formed in these collisions.  To try to gain more insight into this question, collisions of light ions such as oxygen, neon, and magnesium, have been proposed as a way to further examine the relation between ``small" collision systems and ``large" ones~\cite{Brewer:2021kiv,Giacalone:2024ixe,Giacalone:2024luz}.  To this end, oxygen-oxygen (OO) collisions were performed at the Relativistic Heavy Ion Collider (RHIC) in May of 2021 at a center-of-mass collision energy per nucleon pair ($\sqrt{s_{_{\text{NN}}}}$) of 200 GeV~\cite{Huang:2023viw}, and the Large Hadron Collider (LHC) performed OO and neon-neon (NeNe) collisions at $\sqrt{s_{_{\text{NN}}}}=5.36$ TeV in July of 2025.  The availability of these datasets and the first results coming from them~\cite{CMS:2025bta, CMS:2025tga, ATLAS:2025nnt, ALICE:2025luc, CMS-PAS-HIN-25-010, CMS:2026qef} have prompted intense interest in light ion collisions and, by extension, any experimental backgrounds that could affect measurements attempting to gain insight into potential QGP-like effects that manifest in these collisions.  Looking towards the future, light ion collisions have been proposed as a potential direction for the LHC ion physics program beyond the Run 4 era~\cite{Waagaard:2025dsi,Citron:2018lsq}, further motivating the need to study these backgrounds.

Ion beams at a particle collider are subject to many effects that can cause beam losses which are not present for beams of more fundamental particles. As an example, for $^{208}\text{Pb}$ beams at the LHC, the release of a neutron via the process $^{208}\text{Pb} \to {}^{207}\text{Pb}+n$ causes the ion's charge to mass ratio, which to very good approximation is equal to the atomic number $Z$ divided by the mass number $A$ when nuclear binding energies are neglected, to change.  The daughter ion then has an incorrect magnetic rigidity for the LHC beam optics and is unable to continue to circulate in the collider.  This process accounts for a sizable fraction of $\text{Pb}$ beam losses at the LHC~\cite{Bruce:2009bg}.

When encountering the strong electromagnetic fields caused by other ions, light ions are subject to electromagnetic dissociation processes~\cite{Svetlichnyi:2023nim,Pshenichnov:1997un} which can result in the ion breaking up.  Examples of relevant processes for $^{16}\text{O}$ ions include $^{16}\text{O} \to{}^{14}\text{N}+\text{d}$, $^{16}\text{O} \to{}^{12}\text{C}+^{4}\text{He}$,  and $^{16}\text{O} \to{}^{10}\text{B}+{}^{6}\text{Li}$.  Unlike the scenario for heavy ions, the daughters of all of these processes have charge to mass ratios $Z/A=0.5$ that are identical to that of $^{16}\text{O}$ itself and are therefore able to continue to circulate indefinitely in the collider alongside the oxygen.  This may result in beam contamination that grows over time as daughter ion species build up.  For oxygen and neon, it is expected that $^{4}\text{He}$ will be the most likely contaminant due to its strong binding energy and the potential tendency for the nuclear structure of the original ion to contain $\alpha$ clusters~\cite{Summerfield:2021oex,YuanyuanWang:2024sgp}.  Furthermore, neon may be more susceptible to dissociation, and therefore contamination effects, owing to its highly deformed nuclear structure~\cite{Giacalone:2024luz}. Electromagnetic dissociation is expected to occur most often at locations where ion beams are highly focused, i.e., at the collider's interaction points where experiments sit, and should scale with the intensity of the beams at these locations.  Thus, beam contamination effects may become a barrier preventing higher instantaneous luminosities in the future.

From an experimental standpoint, the presence of beam contaminants is highly problematic.  Contaminants result in asymmetric collisions such as (using $^{16}\text{O}$ as an example again) deuteron-oxygen, helium-oxygen, boron-oxygen, carbon-oxygen, and nitrogen-oxygen, as well as less common contaminant-contaminant collisions like helium-helium, helium-carbon or deuteron-nitrogen collisions.  All of these collisions have a different ``size" and can introduce complications into the interpretation of data that is intended to elucidate the dependence of QGP-like effects on collision size.  Furthermore, because experimenters have no control over the the collision impact parameter, a glancing oxygen-oxygen collision may be experimentally indistinguishable from a head-on helium-oxygen collision.  For all of these considerations, it becomes crucial to have robust estimates of the level of contamination that may be present in an ion beam at a given point in time.  Unfortunately the magnitude of this effect is subject to many complicated variables: the details of the collider's beam optics, the exact decay kinematics of the electromagnetic dissociation process, as well as the cross sections for these processes, and is therefore difficult to simulate from first principles.  

In this article, we propose a simple method for experimenters to get a data-driven estimation of the level of beam contamination coming from transmutation processes.  It uses the expected time dependence and collision-system-size dependence of contaminant collisions to define background-free control regions that can be used to track how contamination builds up over time.  This proposed method operates at an ensemble level, and is not able to distinguish between signal and contaminant collisions on an event-by-event basis.   In Sec.~\ref{sec:method} we provide a brief conceptual overview highlighting how the method works.  In Sec.~\ref{sec:sim} we show a simulation study using a simple two-component toy model that illustrates that the method works well under simplifying assumptions.  In Sec.~\ref{sec:caveats} we discuss potential failure modes and weaknesses of the method and how they might be mitigated before briefly summarizing our findings in Sec.~\ref{sec:conclusion}.  

\section{Method overview}
\label{sec:method}

\begin{figure}[htbp]
    \centering
    \includegraphics[width=0.99\columnwidth]{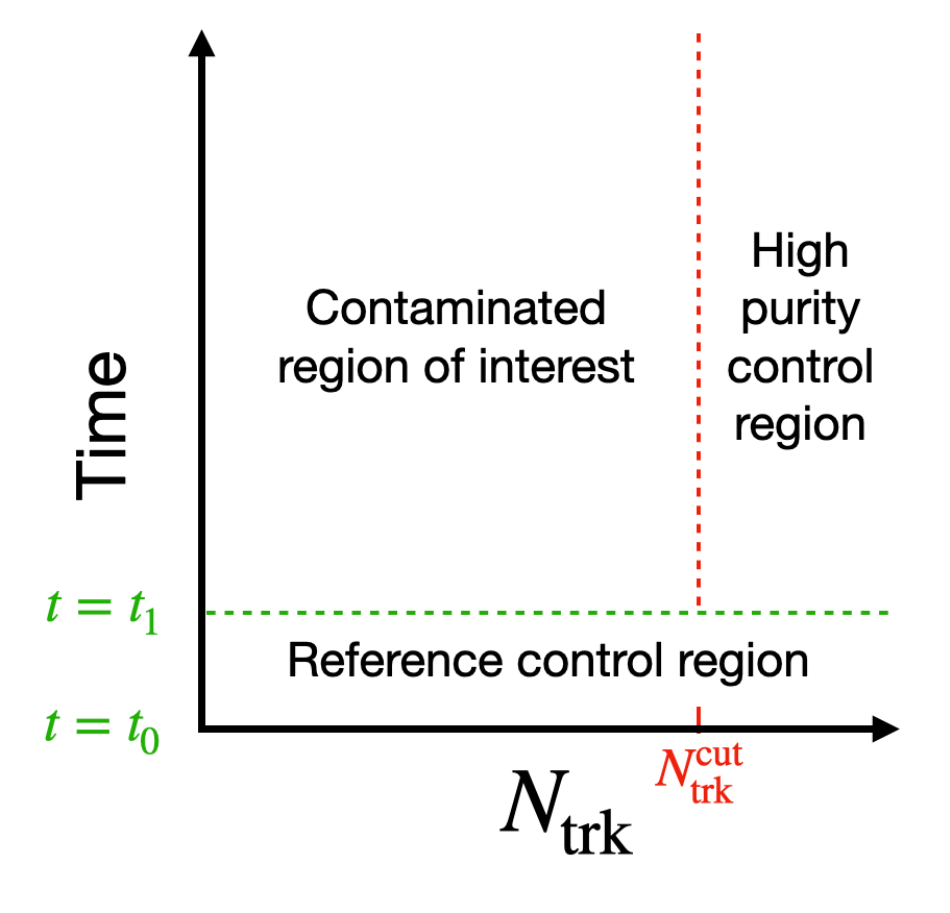}
    \caption{A schematic illustration showing the various regions in the $N_\text{trk}$ versus time plane defined for the method proposed in this work.  Two control regions where contamination is expected to be negligible are defined in order to access potential contamination in the region of interest.}
    \label{fig:illustration}
\end{figure}

At the LHC, a ``fill'' refers specifically to the period between when beams are injected into the collider until when the beams are dumped. During the course of the fill, there are two kinds of effects come into play: the rate of collisions for the injected ions decreases over time as the beam intensity decreases while the transmutation background increases with time. The goal of this method is to estimate the rate of contaminant collisions versus time. 

The problem can be tackled by using an approach essentially equivalent to the ABCD method detailed in Refs.~\cite{Choi:2019mip,Kasieczka:2020pil}.  In that method, two uncorrelated variables are utilized to define control regions to constrain backgrounds in order to search for signal in an adjacent region.  Here, we utilize the expected time dependence of the transmutation effect, alongside with another variable that is correlated with the ``size" of the collision system as our two variables of interest.  This second variable should be correlated with the number of interacting nucleons present in the collision, such that more ``head-on" collisions, or collisions of heavier ions, produce larger values.  Examples of appropriate and common variables are the total number of tracks measured in the event $N_{\text{trk}}$ or the total energy observed in an experiment's calorimeter system.  For convenience of explanation, in the discussion that follows we choose to focus on using $N_{\text{trk}}$ as the second variable of choice.  The proposed method utilizes the two-dimensional plane shown in Fig.~\ref{fig:illustration}, where the $N_{\text{trk}}$ distribution is plotted against the amount of time that has elapsed in a fill.

 At the start of the fill, during a short a time interval that can be defined to be from times $t_0$ to $t_1$, the contribution from beam transmutation is expected to be small because there has not been sufficient time for contaminants to build up.  We label this time as the reference control region in Fig.~\ref{fig:illustration}. The reference control region can be used to establish the uncontaminated shape of the $N_{\text{trk}}$ distribution.  Crucially, the shape of this distribution \textit{will not change} versus fill time for non-contaminant collisions.  However, one cannot simply subtract this uncontaminated distribution from distributions measured at times $t>t_1$, as the total rate of non-contaminated collisions is expected to be smaller for $t>t_1$ because of beam intensity decay.  
 
 This challenge can be overcome by defining a second control region that is used to track the change in the overall normalization of the uncontaminated $N_{\text{trk}}$ distribution versus fill time.  The second control region, which we call the high purity control region can be defined corresponding to \Ntrk values above some chosen cut value \Ntrkcut.  The value of \Ntrkcut should be chosen such that no contaminant collisions (which always have less colliding nucleons than maximally head-on collisions of the initial ions in the beam) produce \Ntrk above \Ntrkcut.  This initial choice for \Ntrkcut can be chosen using simulations of contaminant collisions and refined based on what is seen in data.

A scaling factor, $f$, can be defined to account for the change in the overall scale of the non-contaminant distribution in a later time period $t_1<t<t_2$ as compared to the reference control region.  It compares the uncontaminated tails of the \Ntrk distributions above \Ntrkcut:
\begin{equation}
    f = \frac{\int_{\Ntrkcut}^{\infty} \mathrm{\Ntrk(t_1<t<t_2)~d\Ntrk}}{\int_{\Ntrkcut}^{\infty} \mathrm{\Ntrk(t_0<t<t_1)~d\Ntrk}}.
\end{equation}

 Using this definition, one can extract the time-dependent $\Ntrk$ distribution of the contaminant collisions, $\Ntrk^{\text{cont.}}$, in the contaminated region of interest (which is defined as the space having $t>t_1$ and $\Ntrk<\Ntrkcut$, as seen in Fig.~\ref{fig:illustration}) by simply subtracting the scaled reference \Ntrk distribution:
\begin{multline}
     \Ntrk^{\text{cont.}}(t_1<t<t_2) = \\ \Ntrk(t_1<t<t_2) - f\Ntrk(t_0<t<t_1).
\end{multline}

 The limits of the range $t_1<t<t_2$ can be chosen at will to explore the entire time dependence of contamination effects.  By integrating the various extracted \Ntrk distributions, one can then estimate the fraction or rate of contaminant and non-contaminate collisions at any time $t$ in the fill.


\section{Simulation studies}
\label{sec:sim}

To test the validity of the proposed approach, a simple toy model study was performed using simulated events.  In this model, beams of $^{16}\text{O}^{16}\text{O}$ collisions with only one contaminant species ($^{4}\text{He}$) are considered.  Furthermore, contaminant-contaminant collisions ($^{4}\text{He}^{4}\text{He}$) are neglected, as they will be rare as compared to $^{4}\text{He}^{16}\text{O}$ collisions.  Previously performed accelerator simulations for $^{16}\text{O}$ beams indicate that this approximation of only considering $^{4}\text{He}$ should capture the majority of the beam transmutation effects at the LHC~\cite{transmutation}.

\begin{figure}[tbp]
    \centering
    \includegraphics[width=0.99\columnwidth]{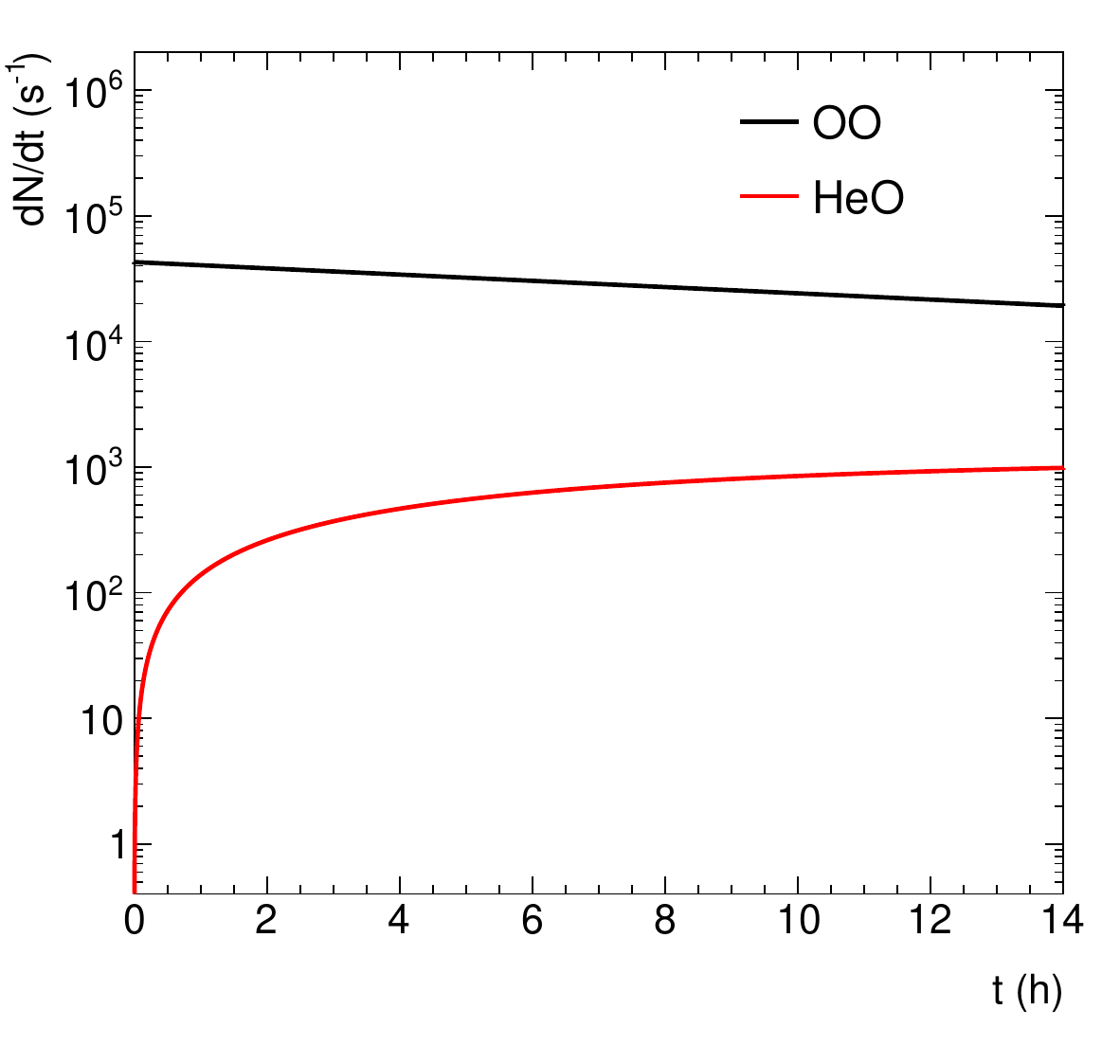}
    \caption{The event rate, $\mathrm{d}N/\mathrm{d}t$, as a function of time up to 14 hours for OO and HeO collisions.  The shapes of these distributions were extracted from Ref.~\cite{transmutation}.}
    \label{fig:Lumi}
\end{figure}

To try to have the model more accurately reflect a realistic LHC fill,  the time-dependent shape of the instantaneous $^{16}\text{O}^{16}\text{O}$ and $^{4}\text{He}^{16}\text{O}$ luminosity is extracted by performing a fit to the luminosity projections in Ref~\cite{transmutation}. Specifically, the time dependence is modeled with simple exponential functions: an exponential decay for OO collisions,
\begin{equation}
    \frac{dN_{\mathrm{OO}}}{dt} = a \, \exp\!\left(-\frac{t}{\tau_{\mathrm{OO}}}\right),
\end{equation}
where $\tau_\mathrm{OO} = 17.5$ hours, and a time-dependent growth of HeO collisions,
\begin{equation}
    \frac{dN_{\mathrm{HeO}}}{dt} = b \, \left[1 - \exp\!\left(-\frac{t}{\tau_{\mathrm{HeO}}}\right)\right],
\end{equation}
where $\tau_\mathrm{HeO} = 8.1$ hours. Figure~\ref{fig:Lumi} shows the time evolution of the resulting event rates, $\mathrm{d}N/\mathrm{d}t$, up to 14 hours.  A hadronic cross section of 1.3 b for OO collisions was used to convert the extracted instantaneous luminosity to an event rate, which defines the overall scale factor $a$.  The event rate of contaminant HeO collisions, set by the $b$ parameter, is scaled to give an arbitrary 5\% HeO contamination level by the end of a 14-hour fill. Note that the absolutes rates of OO and HeO collisions do not matter for the study; only their relative contributions matter for the purposes of illustrating the proposed technique.

\begin{figure*}[htbp]
    \centering
    \includegraphics[width=1.0\linewidth]{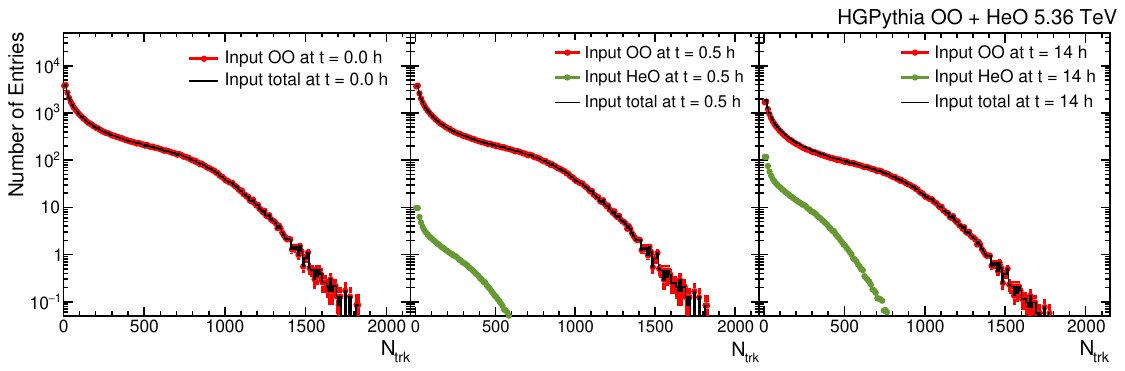}
    \caption{The input \Ntrk distributions for OO and HeO events at $t=0$ hours (left), $t=0.5$ hours (middle) and $t=14$ hours (right), generated using HG-Pythia.}
    \label{fig:NtrkInput}
\end{figure*}

\begin{figure*}[htbp]
    \centering
    \includegraphics[width=1.0\linewidth]{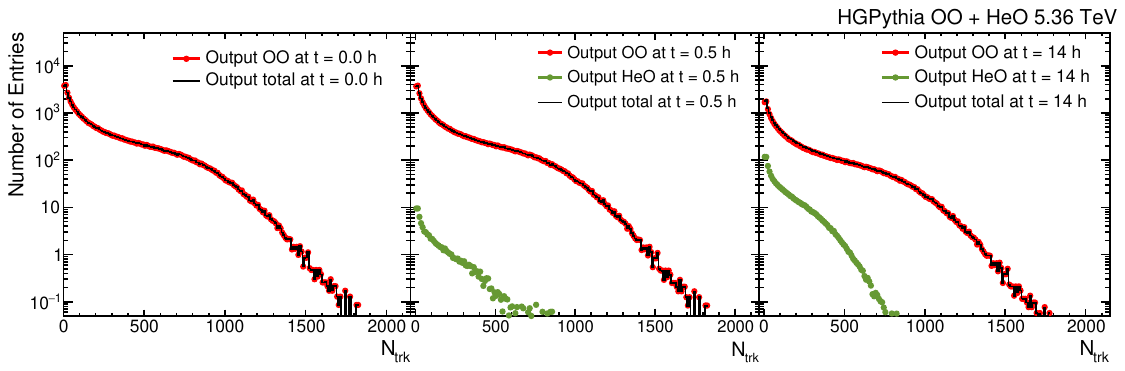}
    \caption{The extracted OO and HeO components \Ntrk distributions along with the total distribution for OO and HeO events at $t=0.0$ hours (left), $t=0.5$ hours (middle) and $t=14$ hours (right), generated using HG-Pythia.}
    \label{fig:NtrkOutput}
\end{figure*}

To model the \Ntrk distributions for these collisions, we use the HG-Pythia model~\cite{Loizides:2017sqq}, a simple toy model that uses Glauber model for the initial state and Pythia for the final state. For a given MC Glauber event, PYTHIA (v6.28 with Perugia 2011 tune) is run to generate pp events with exactly ${{(N^\mathrm{hard}_\mathrm{NN}})}_i$ multiple parton interactions for each nucleon-nucleon collision $i$, where $i$ runs from 0 to the total number of nucleon-nucleon collisions, $N_\mathrm{coll}$. The generated particles from all PYTHIA events are then combined and treated like a single ion collision in the further analysis. HG-Pythia is not intended to necessarily correctly reproduce the overall scale of the OO \Ntrk distribution, but it does give a reasonable qualitative shape that can be used to illustrate the salient points of our proposed method. We generate both OO and HeO collisions at $\sqrt{s_\mathrm{_\text{NN}}}$ = 5.36 TeV. 

Figure~\ref{fig:NtrkInput} shows the input \Ntrk distributions used for three time slices, t=0 hours (left), t = 0.5 hours (middle) and t = 14 hours (right). These distributions are scaled to match the respective event rates $\mathrm{d}N/\mathrm{d}t$ from Fig.~\ref{fig:Lumi}. The \Ntrk distribution for OO is shown in red markers and HeO is shown in green markers.  The sum of these distributions (which is what would actually be measured by an experiment) is shown by the black line. The time dependence of both of these distributions is evident after comparing the three panels: the OO collisions decay over time and the HeO collisions grow with time.  This leads to a subtle change in shape of the black line with time.  Importantly, the input HeO distribution contributes significantly only in the low-\Ntrk region, approximately below $\Ntrk = 900$. Because of this, \Ntrkcut is chosen to be $900$ for this implementation of our proposed method.

Figure~\ref{fig:NtrkOutput} shows the results after applying our contamination extraction method.  The black lines are the same as in Fig.~\ref{fig:NtrkInput}.  From these total distributions, the OO and HeO contributions are extracted.  The left panel shows the measurement in the reference control region, where the total is defined to be entirely resulting from OO collisions.  The middle and right panels show that, when scaling the OO distribution using the high purity control region (illustrated by the red markers), the difference between the total and the OO estimation gives a distribution for the HeO contribution (green markers).  The shapes of the output HeO \Ntrk distributions qualitatively match the shapes of the input distributions shown in Fig.~\ref{fig:NtrkInput}.  Note that small fluctuations in the range of $\Ntrk\sim\Ntrkcut$ are the result of statistical fluctuations when sampling the high \Ntrk tails of the distributions.

\begin{figure}[tbp]
    \centering
    \includegraphics[width=0.99\columnwidth]{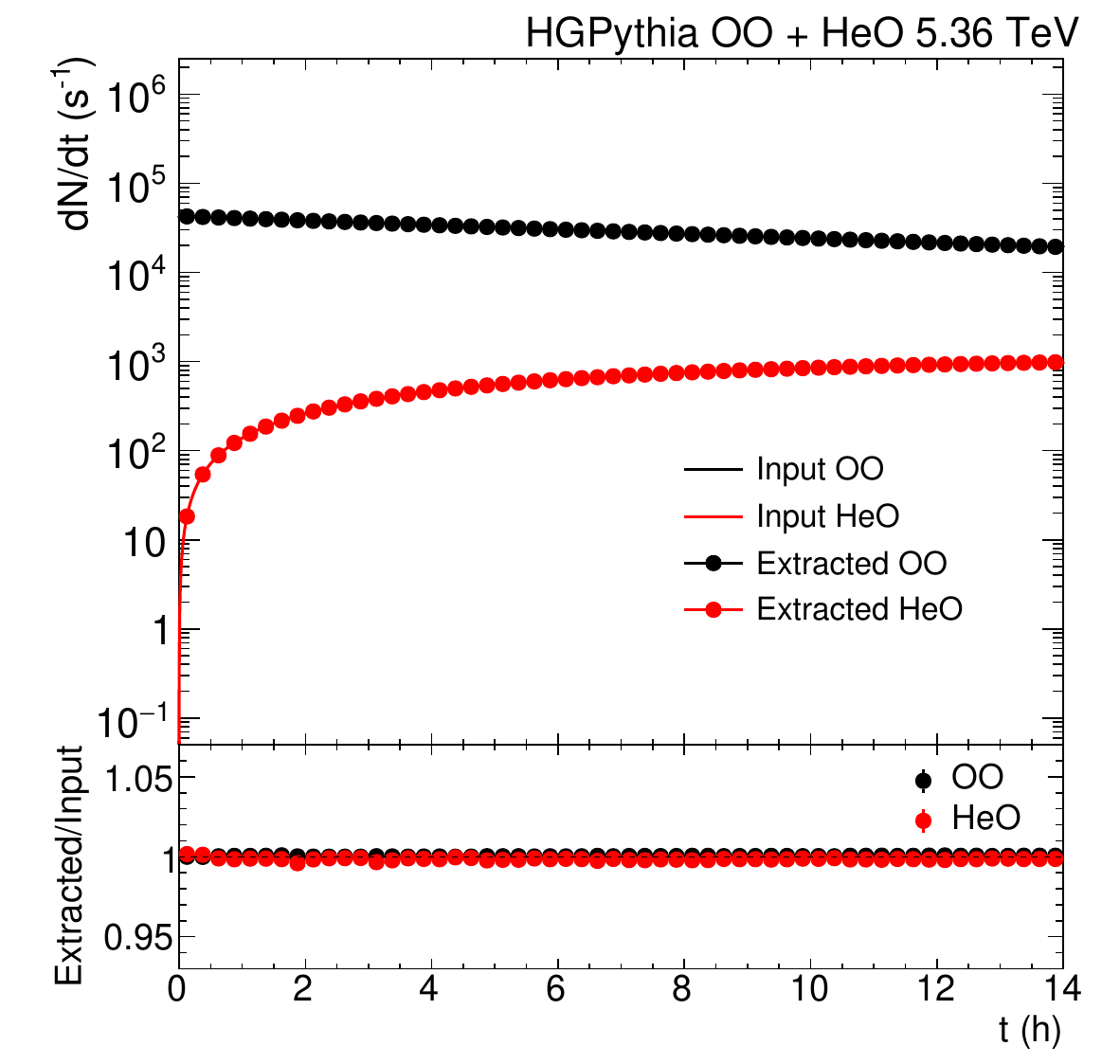}
    \caption{Closure test of the \Ntrk-based separation method. The OO and HeO components (circular markers) are extracted from the total \Ntrk distributions and compared to the input collision rates (lines), showing very good agreement.}
    \label{fig:NtrkClosure}
\end{figure}

The validity of the method is assessed through a closure test, as shown in Fig.~\ref{fig:NtrkClosure}.  The closure is tested by comparing the extracted OO and HeO rates, calculated by integrating the distributions in Fig.~\ref{fig:NtrkOutput}, against the original inputs. The sub-percent-level agreement observed across the fill time demonstrates the ability of the method to provide reliable contamination estimates in this toy model.  





\begin{figure*}[htbp]
    \centering
    \includegraphics[width=0.49\linewidth]{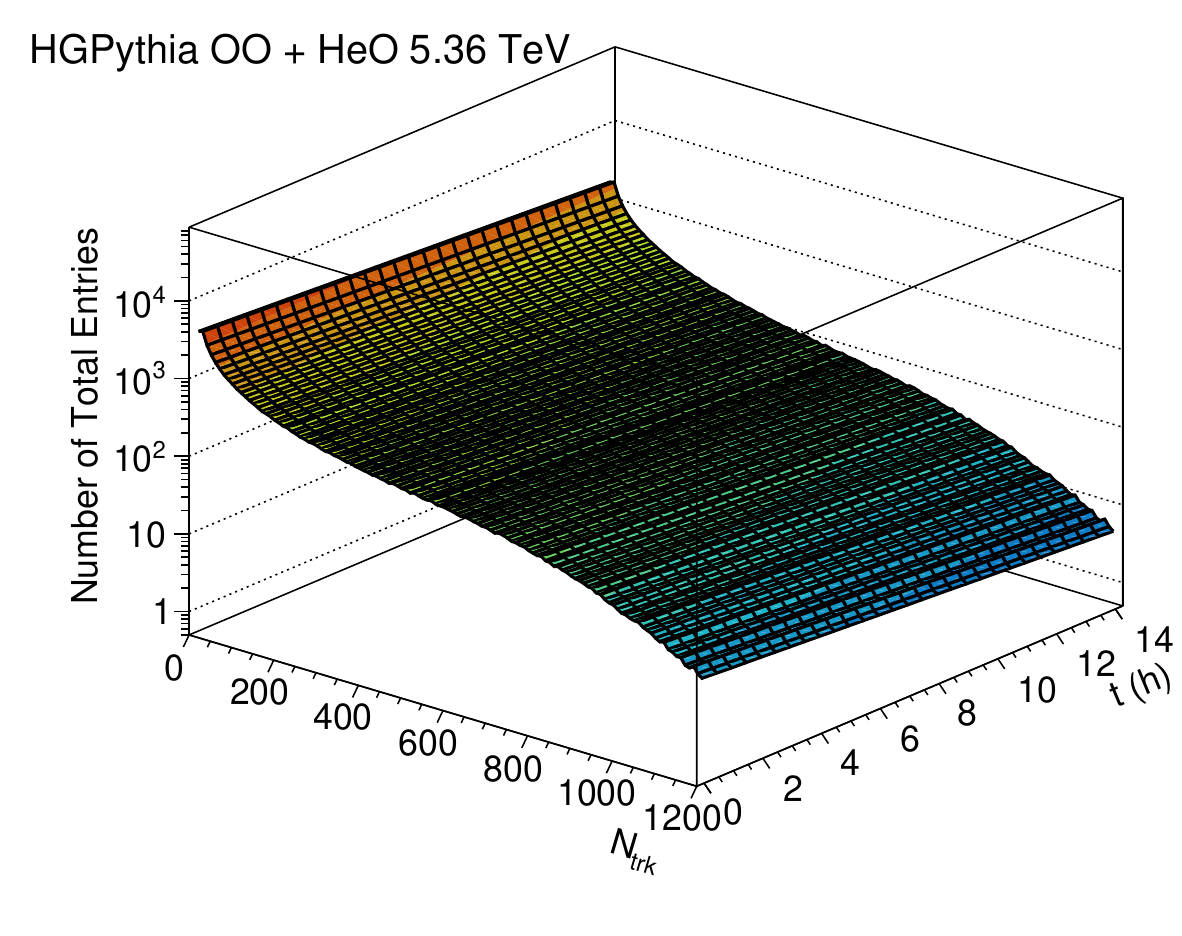}
    \includegraphics[width=0.49\linewidth]{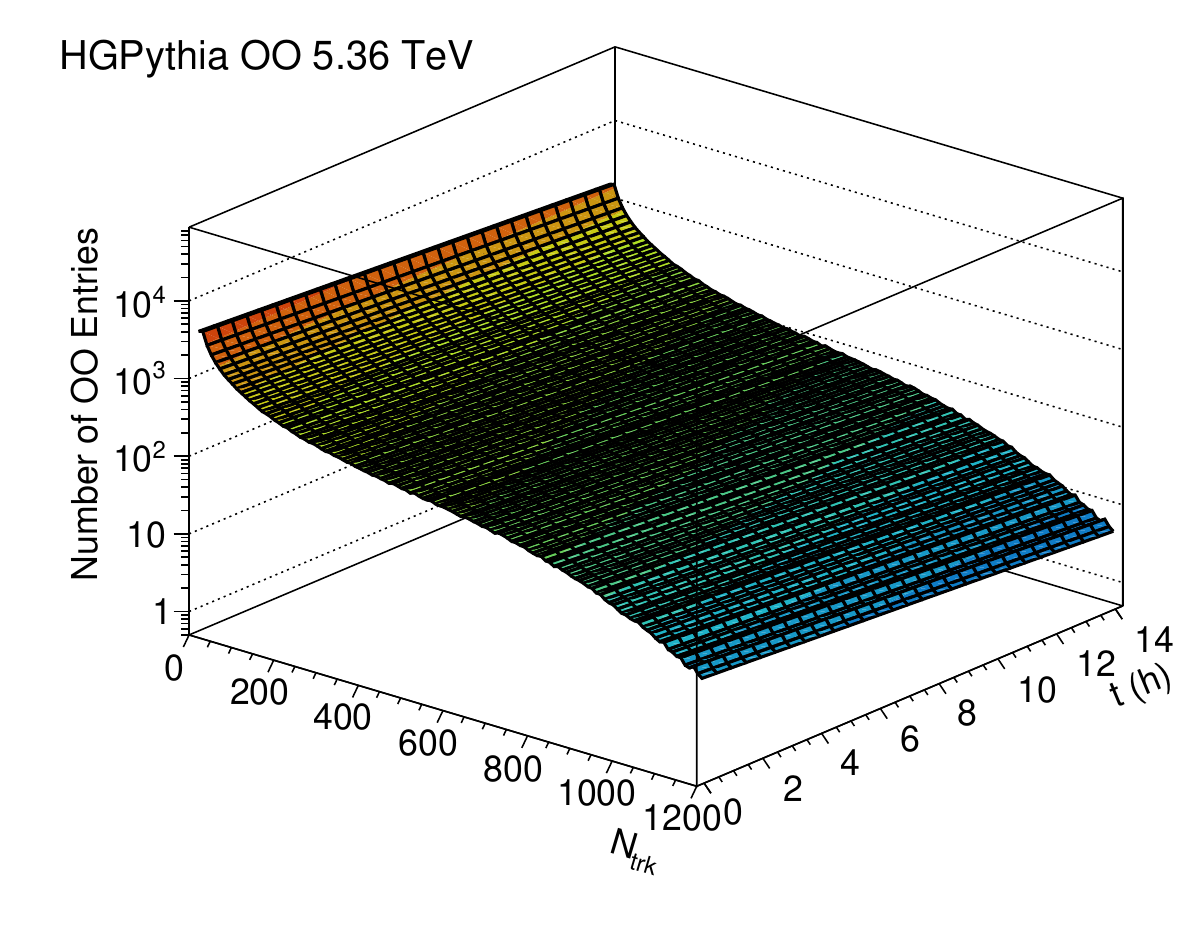}
    \includegraphics[width=0.49\linewidth]{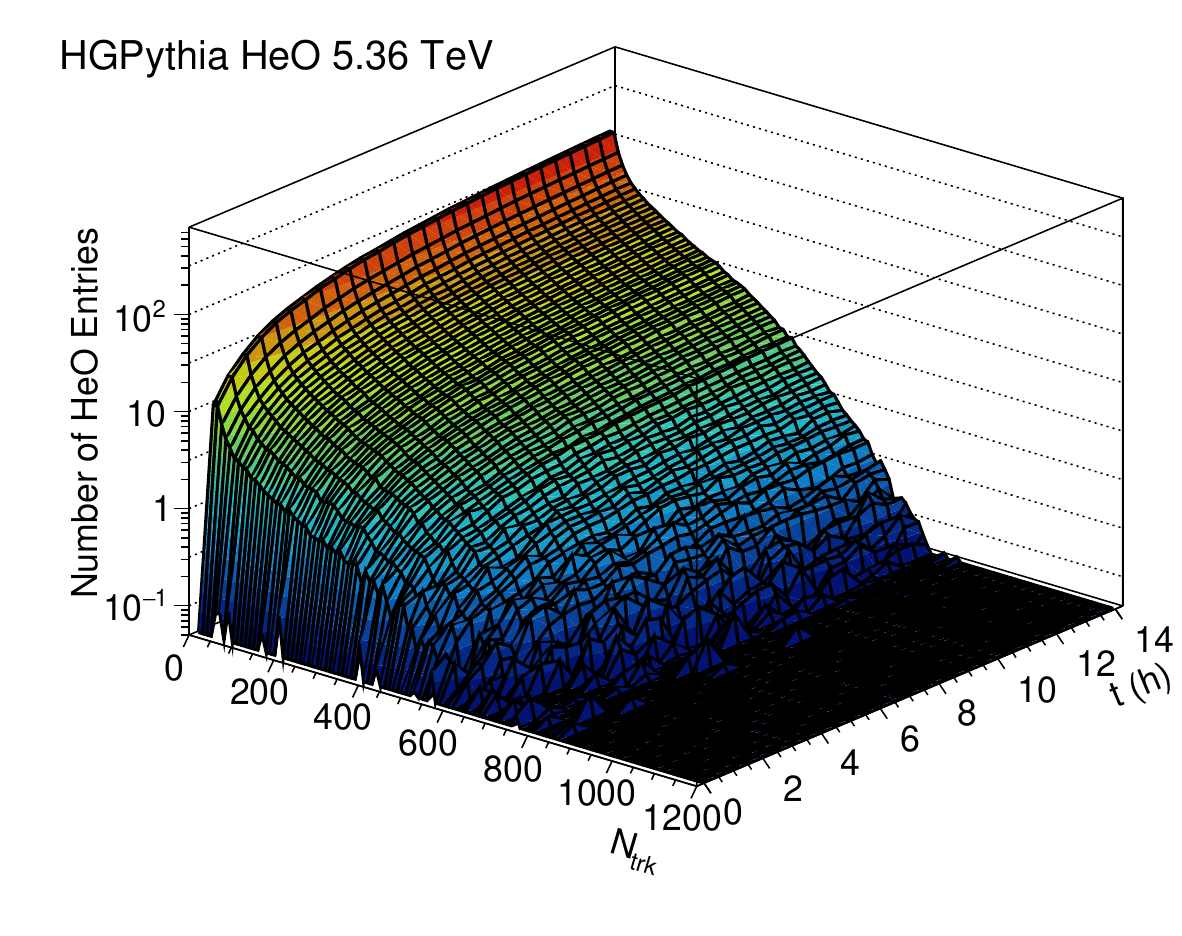}
    \caption{Time evolution of the total (top left) \Ntrk distributions, as well as the extracted OO (top right) and HeO (bottom) \Ntrk distributions after applying the proposed procedure. The HeO contribution grows over time, predominantly in the low-\Ntrk region.}
    \label{fig:NtrkTime}
\end{figure*}

For illustrative purposes, only three time slices were displayed in the previous figures.  However, one can perform the same procedure for many time slices during a fill to get an idea of the overall evolution and structure of the $\Ntrk$ distribution. Figure~\ref{fig:NtrkTime} shows the time evolution of the total \Ntrk distribution and the separated OO and HeO components, across the full duration of a fill. As time progresses, the contribution from HeO events becomes increasingly visible, especially in the low-\Ntrk region (up to approximately \Ntrk = 900), reflecting the gradual buildup of HeO contamination.  


\section{Complicating factors}
\label{sec:caveats}

The studies presented above make some simplifying assumptions that are not realistic in the context of a real experimental environment.  Experimentally, the effect of these assumptions could introduce uncertainties into the extraction of the value of beam contamination.  Fortunately, the high level of precision that modern collider experiments can achieve should allow many of these effects to be mitigated with careful study.  In some cases mitigation strategies require specific beam conditions, highlighting the need for experimenters to have a well-defined plan for dealing with contamination effects before data is collected.  Here, we discuss three major caveats that are relevant for the proposed method of contamination estimation, and how they can be mitigated.

\subsection{Pileup}

Modern colliders have the ability to deliver high instantaneous luminosities.  This, when combined with the relatively large hadronic cross section for ion-ion collisions (on the order of barns), leads to a substantial probability for multiple interactions to occur during the same bunch crossing.  This phenomenon is typically referred to as ``pileup."  

Pileup effectively injects additional particle activity into an event, causing a migration from a ``true" value of event activity to a larger observed value. This effect can distort the distribution of the event activity variable (such as $N_{\mathrm{trk}}$) used in the contamination extraction method.  Since pileup is proportional to the instantaneous luminosity, which is usually time dependent itself, this can cause a time-dependent change in the $N_{\mathrm{trk}}$ distribution.  The contamination extraction method relies on an assumption that the shape of the $N_{\mathrm{trk}}$ distribution measured in the reference control region is constant versus time if there are no contaminants.  Pileup effects clearly violate this assumption and can result in over-subtraction of the scaled OO template from the contaminated region of interest.  This leads to an underestimation of the contamination rate.  

Pileup has been a major experimental challenge for the high-energy physics community at the LHC for over a decade, and therefore many experiments have well-defined and robust pileup mitigation strategies.  After data with pileup is collected, modern collider experiments typically leverage precise particle tracking to identify the presence of multiple interaction vertices.  This enables tagging of pileup events.  Vetoing on pileup events when performing the contamination extraction is one potential strategy that could be used to deal with pileup effects, although careful consideration must be taken to ensure that any veto conditions do not introduce distortions of the $N_\mathrm{trk}$ distribution itself.  In principle, such studies of veto conditions can could be performed using simulations.  As an alternative approach, if a reliable association can be made between the particles measured in the experiment and the interaction vertex from which they originate, each individual vertex in a bunch crossing can be analyzed as an independent event for the purpose of extracting contamination effects.  Once again though, this method also relies upon a strong understanding of the interaction vertex reconstruction performance and track-vertex association fidelity associated with the experiment doing the measurement.

\begin{figure*}[htbp]
    \centering
    \includegraphics[width=0.95\linewidth]{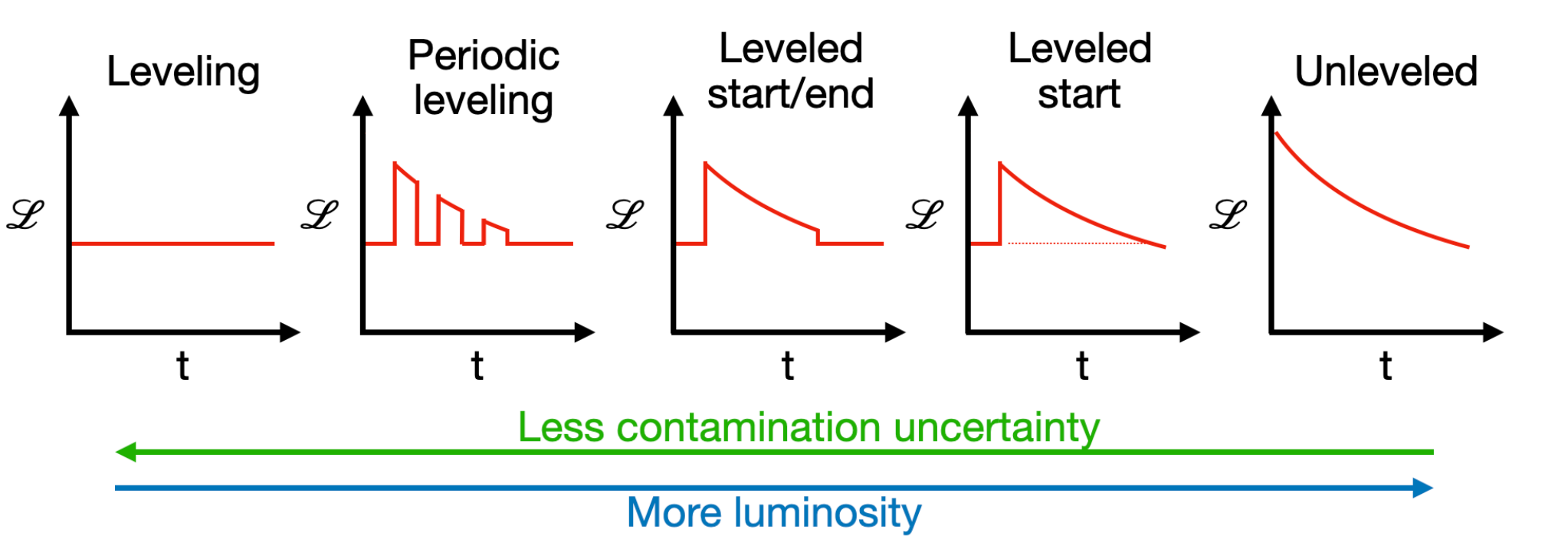}
    \caption{Schematic representations of various strategies that can be used to mitigate pileup effects when extracting beam contamination rates.  Strategies on the left are expected to give more precise contamination results but limit the sampled luminosity, whereas strategies on the right maximize the luminosity at the cost of more pronounced pileup contributions.}
    \label{fig:levelingstrats}
\end{figure*}

Another method for dealing with pileup effects involves careful planning of the beam conditions when collecting the dataset to be analyzed.  Modern colliders are able to add a small spatial offset to colliding particle beams, which effectively reduces the instantaneous luminosity delivered to an interaction point.  This beam separation process can be used manipulate the luminosity profile (and therefore the pileup profile) seen by an experiment.  In particular, the instantaneous luminosity can be ``leveled" by separating the beams to achieve a desired pileup value and then slowly bringing the beams back together over time to compensate for decreasing beam intensities.  Requesting leveling at a fixed value for the duration of a fill is a natural way that an experiment can remove the time-dependent pileup effects that could confound contamination extraction.

The leveling process does come with a downside: the separation of the beams reduces the integrated luminosity sampled by an experiment, potentially hindering luminosity-hungry physics programs that are reliant on very rare processes.  It is worth noting that a number of ``hybrid" leveling strategies could be pursued to reach a reasonable compromise between the sampled integrated luminosity and the level of precision that beam contamination can be extracted.  A schematic sketch of these strategies, shown as a plot of the collected instantaneous luminosity vs time, is shown in Fig.~\ref{fig:levelingstrats}.  The most conservative strategy is to fully level for the duration of the fill, which effectively removes the pileup uncertainty from this method.  A more aggressive strategy would be to request leveling at the beginning of a fill to establish the reference control region and then periodically return to the leveled value to establish measurements of the contamination versus time.  These measurements can then be interpolated to estimate the full time dependence of the contamination effect for the entire duration of the fill.  Alternatively, just the beginning and the end of the fill can be leveled to establish a measurement of the contamination at the end of the fill, where it is expected to be largest.  This strategy could be appropriate if only an upper limit on the amount of contamination is needed.  Perhaps the most aggressive leveling strategy would be leveling at the beginning of the fill to a value that the luminosity is expected to naturally decay to at the end of the fill.  This would be essentially equivalent to leveling at both the beginning and the end of the fill without actually realizing any luminosity losses at the end of the fill.


\subsection{Multiple contaminants}

For light ion beams having $Z/A = 0.5$, many potential contaminant ions are able to continue to circulate in the collider.  To illustrate our method of estimating transmutation backgrounds, we made a simplifying assumption that there is one dominant contaminant species ($^{4}\text{He}$ in the case of $^{16}\text{O}$ beams).  The presence of more than one contaminant species potentially complicates the structure of the contaminant $N_{\text{trk}}$ distribution, as it is would no longer be expected to have only one clear ``knee" above which the distribution quickly falls to zero.  The method we propose does not depend on the specific shape of the overall contaminant $N_{\text{trk}}$ distribution (and in fact it could be used to extract the shape of this distribution as shown in the bottom panel of Fig.~\ref{fig:NtrkTime}) as long as the value of $N_{\text{trk}}^{\text{cut}}$ is selected such that all contaminant events produce less tracks than the chosen threshold.  However, additional tails coming from heavier contaminant collisions, such as $^{14}\text{N}^{16}\text{O}$ in the case of $^{16}\text{O}$ beams will have a knee structure that is very close to the upper $N_{\text{trk}}$ limit for $^{16}\text{O}^{16}\text{O}$, and therefore could potentially bleed into the high purity control region if the $N_{\text{trk}}^{\text{cut}}$ value is too low.  To evaluate the impact of this potential effect, an upward scan of the experimenter's nominal $N_{\text{trk}}^{\text{cut}}$ choice can be performed to search for additional ``knee" structures in the extracted contaminant distribution.  A lack of observable structures above the nominal $N_{\text{trk}}^{\text{cut}}$ choice would imply that all of the relevant contaminants are accounted for by the current choice, whereas the presence of additional structures would suggest that the $N_{\text{trk}}^{\text{cut}}$ value needs to be raised to account for additional contaminants.  In this way, the experimenter can systematically explore the degree to which additional contaminants may be relevant.  Statistical considerations for the event yield in the high-purity control region are expected to limit the maximal value that \Ntrkcut can take, so experiments gathering a high rate of data may be able to be more conservative with their choice of \Ntrkcut.

\begin{figure}[bt]
    \centering
    \includegraphics[width=0.99\linewidth]{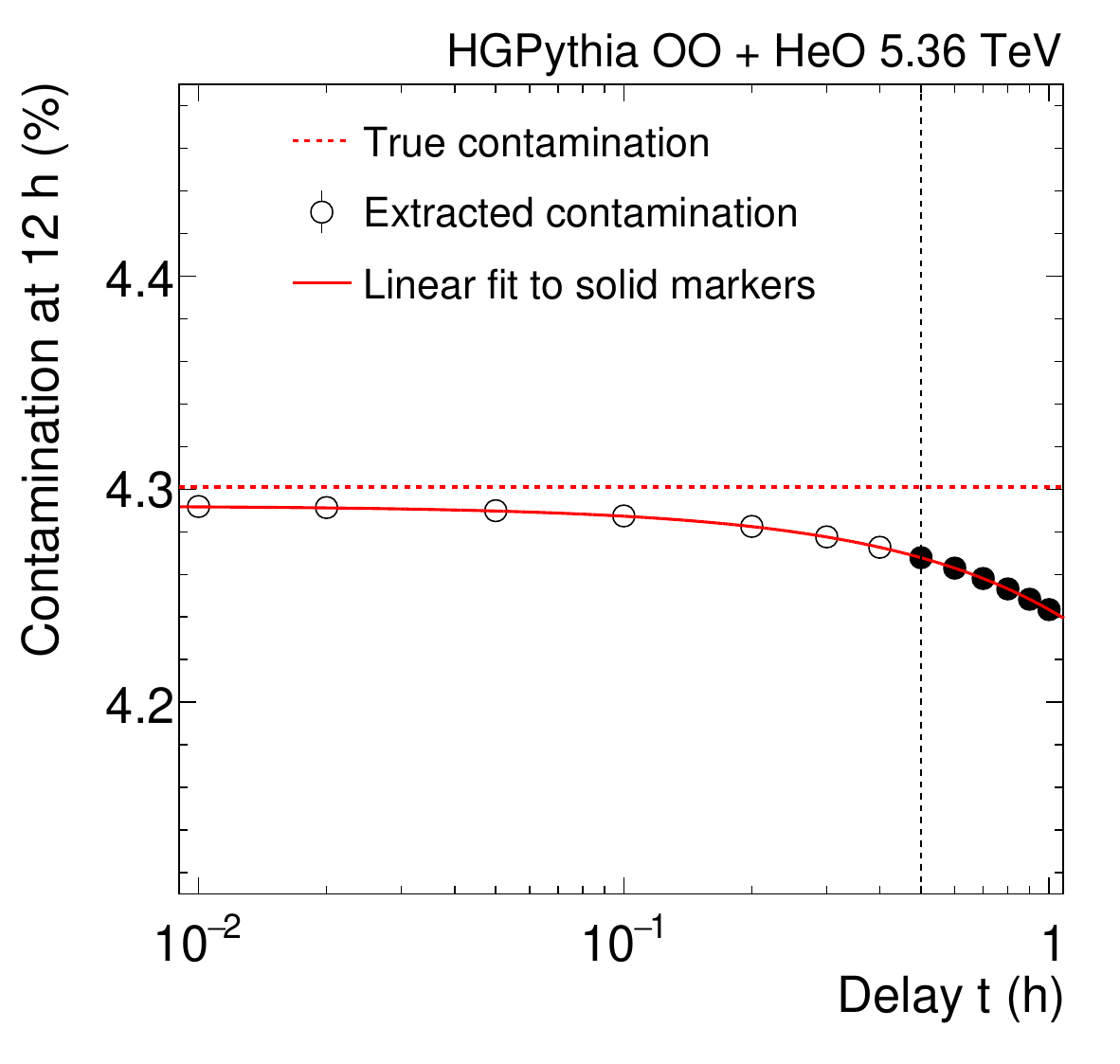}
    \caption{Percentage-level contamination at 12 hours as a function of the delayed time. The vertical black dashed line indicates the start of data collection at $t = 0.5~\text{h}$. The solid red line shows a linear fit to the data points with $t \geq 0.5~\text{h}$ (solid black markers), while the open markers denote data points at $t < 0.5$ h that are not included in the fit. The horizontal red dotted line denotes the “true’’ contamination at $t = 0$, which is within 0.01\% of the contamination extracted in the limit $t \to 0$, e.g., at $t = 0.01$~h.}
    \label{fig:delayedstart}
\end{figure}


\subsection{Delayed start of data collection}

Even if an ion beam starts its lifetime without any contamination, transmutation will begin essentially as soon as a sufficient instantaneous luminosity is achieved to catalyze a high rate of electromagnetic dissociation.  Experiments are commonly unable to begin data collection immediately after this point because of detector protection requirements---some short period of time (typically less than a few minutes) is usually needed to ramp detector voltages, move non-static detector components to their nominal data-taking parameters, etc.  The result of this delay is that the reference control region will never be truly free of contamination.  The presence of some contamination in this control region will result in an over subtraction in the region of interest, producing an underestimate of the total contamination rate.


The effects of a delayed start of data collection can be estimated by repeating the contamination extraction using several choices of the reference control region, each corresponding to a different delay after the onset of high-intensity collisions. The extracted contamination at a fixed time (e.g., 12 hours) can then be plotted as a function of the delay to study how it varies with the start time. In Fig.~\ref{fig:delayedstart}, the vertical dashed line at $t = 0.5$~h marks the actual start of data collection. A linear fit is applied to the data for $t \geq 0.5$~h and extrapolated to earlier times to infer the true contamination at the onset of collisions. The extrapolation function is not guaranteed to be linear, and the experimenter should choose a reasonable functional form based on the data or accelerator simulations. The extrapolated value at an earlier time (eg: $t = 0.01$~h) is compared to the true contamination calculated from Fig.~\ref{fig:Lumi}, showing excellent agreement within 0.01\%.


\subsection{A word of caution}

Given that the contaminant $N_{\text{trk}}$ distribution can be extracted as a function of time with this method, as shown in the bottom panel of Fig.~\ref{fig:NtrkTime}, it is tempting to use this template to subtract contamination effects and get a new measurement of $\Ntrk$ for collisions of the injected ions.  However, the resulting $N_{\text{trk}}$ distribution will be equal to a scaled version of the reference control region \textit{by definition}, meaning that the result in the region of interest would not constitute an independent measurement.  Thus, this method is appropriate for determining an estimation of the level of contamination, but careful consideration should be taken before using the output distributions for further manipulation and analysis of data.



\section{Conclusion}
\label{sec:conclusion}
In summary, a data-driven method to estimate potential beam-transmutation contamination rates in light-ion collisions by exploiting their distinct event-occupancy distribution shapes and the time dependence of expected transmutation effects has been proposed. Using HG-Pythia to generate OO and HeO events, a toy model for the total \Ntrk distribution as a function of time for a typical LHC fill was created.  We demonstrate that the method is able to reliably estimate the level of beam contamination under some simplifying assumptions. Potential experimental difficulties with the proposed method are discussed and recommendations for mitigating these challenges are given.  This study is intended to inform discussion of beam transmutation contamination and relevant analysis strategies in recently collected OO and NeNe datasets, as well as in future light-ion runs at the LHC and other ion-ion colliders.  Finally, it is worth noting that a strong understanding of contamination effects may ultimately result in the ability to use contaminant collisions for novel physics analyses, such as the those proposed in Ref.~\cite{Nijs:2025qxm}, but we leave those discussions for future work.

\section{Acknowledgment}

This work is supported by the U.S. Department of Energy, Grant No. DE-FG02-94ER40865.

\section{Data Availability}

The data that support the findings of this article are not publicly available. The data are available from the authors upon reasonable request.





% ...

\bibliography{ref}

@article{Loizides:2017sqq,
    author = "Loizides, Constantin and Morsch, Andreas",
    title = "{Absence of jet quenching in peripheral nucleus{\textendash}nucleus collisions}",
    eprint = "",
    archivePrefix = "arXiv",
    primaryClass = "nucl-ex",
    doi = "10.1016/j.physletb.2017.09.002",
    journal = "Phys. Lett. B",
    volume = "773",
    pages = "408--411",
    year = "2017"
}

@article{Transmutation,
  author  = {Jowett, J. and others},
  title   = {Beam transmutation in the O-O run},
  journal = {LPC Presentation, Indico},
  year    = {2025},
  url     = {https://indico.cern.ch/event/1558885/contributions/6565908/attachments/3086988/5465745/OO%20transmutation-dp1.pdf}
}

@unpublished{CMS:2025bta,
    author = "Hayrapetyan, Aram and others",
    collaboration = "CMS",
    title = "{Discovery of suppressed charged-particle production in ultrarelativistic oxygen-oxygen collisions}",
    eprint = "2510.09864",
    archivePrefix = "arXiv",
    primaryClass = "nucl-ex",
    reportNumber = "CMS-HIN-25-008, CERN-EP-2025-226",
    month = "10",
    year = "2025",
}

@unpublished{CMS:2025tga,
    author = "Hayrapetyan, Aram and others",
    collaboration = "CMS",
    title = "{Observation of long-range collective flow in OO and NeNe collisions and implications for nuclear structure studies}",
    eprint = "2510.02580",
    archivePrefix = "arXiv",
    primaryClass = "nucl-ex",
    reportNumber = "CMS-HIN-25-009, CERN-EP-2025-222",
    month = "10",
    year = "2025",
}

@techreport{CMS-PAS-HIN-25-010,
      collaboration = "CMS",
      title         = "{Pseudorapidity distributions of charged hadrons in oxygen-oxygen collisions at 5.36 TeV}",
      institution   = "CERN",
      reportNumber  = "CMS-PAS-HIN-25-010",
      address       = "Geneva",
      year          = "2025",
      url           = "https://cds.cern.ch/record/2942012",
}

@techreport{CMS:2026qef,
    author = "Belyaev, Andrey and others",
    collaboration = "CMS",
    title = "{System-size dependence of charged-particle suppression in ultrarelativistic nucleus-nucleus collisions}",
    eprint = "2602.21325",
    archivePrefix = "arXiv",
    primaryClass = "nucl-ex",
    reportNumber = "CMS-HIN-25-014",
    month = "2",
    year = "2026"
}

@unpublished{ATLAS:2025nnt,
    author = "Aad, Georges and others",
    collaboration = "ATLAS",
    title = "{Measurement of the azimuthal anisotropy of charged particles in $\sqrt{s_{_\text{NN}}}=5.36$ TeV $^{16}$O$+^{16}$O and $^{20}$Ne$+^{20}$Ne collisions with the ATLAS detector}",
    eprint = "2509.05171",
    archivePrefix = "arXiv",
    primaryClass = "nucl-ex",
    reportNumber = "CERN-EP-2025-200",
    month = "9",
    year = "2025",
}

@article{Waagaard:2025dsi,
    author = "Waagaard, E. and Bruce, R. and Fernandez, R. Alemany and Bartosik, H. and Jowett, J. M. and Triantafyllou, N.",
    title = "{Charting the luminosity capabilities of the CERN Large Hadron Collider with various nuclear species}",
    eprint = "",
    archivePrefix = "arXiv",
    primaryClass = "physics.acc-ph",
    doi = "10.1016/j.nima.2025.171118",
    journal = "Nucl. Instrum. Meth. A",
    volume = "1083",
    pages = "171118",
    year = "2026"
}

@article{Citron:2018lsq,
    author = "Citron, Z. and others",
    editor = "Dainese, Andrea and Mangano, Michelangelo and Meyer, Andreas B. and Nisati, Aleandro and Salam, Gavin and Vesterinen, Mika Anton",
    title = "{Report from Working Group 5}: {Future physics opportunities for high-density QCD at the LHC with heavy-ion and proton beams}",
    eprint = "",
    archivePrefix = "arXiv",
    primaryClass = "hep-ph",
    reportNumber = "CERN-LPCC-2018-07",
    doi = "10.23731/CYRM-2019-007.1159",
    journal = "CERN Yellow Rep. Monogr.",
    volume = "7",
    pages = "1159--1410",
    year = "2019"
}

@unpublished{ALICE:2025luc,
    author = "Abualrob, Ibrahim Jaser and others",
    collaboration = "ALICE",
    title = "{Evidence of nuclear geometry-driven anisotropic flow in OO and Ne$-$Ne collisions at $\mathbf{\sqrt{{\textit s}_{_\text{NN}}}}$ = 5.36 TeV}",
    eprint = "2509.06428",
    archivePrefix = "arXiv",
    primaryClass = "nucl-ex",
    reportNumber = "CERN-EP-2025-203",
    month = "9",
    year = "2025",
}

@article{Bruce:2009bg,
    author = "Bruce, R. and Gilardoni, S. and Jowett, J. M. and Bocian, D.",
    title = "{Beam losses from ultra-peripheral nuclear collisions between (Pb-208)**82+ ions in the Large Hadron Collider and their alleviation}",
    eprint = "",
    archivePrefix = "arXiv",
    primaryClass = "physics.acc-ph",
    reportNumber = "FERMILAB-PUB-09-367-TD",
    doi = "10.1103/PhysRevSTAB.12.071002",
    journal = "Phys. Rev. ST Accel. Beams",
    volume = "12",
    pages = "071002",
    year = "2009"
}

@article{Kasieczka:2020pil,
    author = "Kasieczka, Gregor and Nachman, Benjamin and Schwartz, Matthew D. and Shih, David",
    title = "{Automating the ABCD method with machine learning}",
    eprint = "",
    archivePrefix = "arXiv",
    primaryClass = "hep-ph",
    doi = "10.1103/PhysRevD.103.035021",
    journal = "Phys. Rev. D",
    volume = "103",
    number = "3",
    pages = "035021",
    year = "2021"
}

@article{Pshenichnov:1997un,
    author = "Pshenichnov, I. A. and Mishustin, I. N. and Bondorf, J. P. and Botvina, A. S. and Ilinov, A. S.",
    title = "{Nuclear multifragmentation induced by electromagnetic fields of ultrarelativistic heavy ions}",
    eprint = "",
    archivePrefix = "arXiv",
    doi = "10.1103/PhysRevC.57.1920",
    journal = "Phys. Rev. C",
    volume = "57",
    pages = "1920--1926",
    year = "1998"
}

@article{Svetlichnyi:2023nim,
    author = "Svetlichnyi, Aleksandr and Savenkov, Savva and Nepeivoda, Roman and Pshenichnov, Igor",
    title = "{Clustering in Oxygen Nuclei and Spectator Fragments in $^{16}$O{\textendash}$^{16}$O Collisions at the LHC}",
    doi = "10.3390/physics5020027",
    journal = "MDPI Physics",
    volume = "5",
    number = "2",
    pages = "381--390",
    year = "2023"
}

@article{Nijs:2025qxm,
    author = "Nijs, Govert and van der Schee, Wilke",
    title = "{Transmutation of 16O and 20Ne at the Large Hadron Collider}",
    eprint = "",
    archivePrefix = "arXiv",
    primaryClass = "nucl-th",
    reportNumber = "CERN-TH-2025-127",
    doi = "10.1016/j.physletb.2025.140061",
    journal = "Phys. Lett. B",
    volume = "872",
    pages = "140061",
    year = "2026"
}

@article{Harris:2023tti,
    author = {Harris, John W. and M{\"u}ller, Berndt},
    title = "{``QGP Signatures'' Revisited}",
    eprint = "",
    archivePrefix = "arXiv",
    primaryClass = "hep-ph",
    doi = "10.1140/epjc/s10052-024-12533-y",
    journal = "Eur. Phys. J. C",
    volume = "84",
    number = "3",
    pages = "247",
    year = "2024"
}

@article{Niida:2021wut,
    author = "Niida, T. and Miake, Y.",
    title = "{Signatures of QGP at RHIC and the LHC}",
    eprint = "",
    archivePrefix = "arXiv",
    primaryClass = "nucl-ex",
    doi = "10.1007/s43673-021-00014-3",
    journal = "AAPPS Bull.",
    volume = "31",
    number = "1",
    pages = "12",
    year = "2021"
}

@book{Romatschke:2017ejr,
    author = "Romatschke, Paul and Romatschke, Ulrike",
    title = "{Relativistic Fluid Dynamics In and Out of Equilibrium}",
    eprint = "",
    archivePrefix = "arXiv",
    primaryClass = "nucl-th",
    doi = "10.1017/9781108651998",
    isbn = "978-1-108-48368-1, 978-1-108-75002-8",
    publisher = "Cambridge University Press",
    series = "Cambridge Monographs on Mathematical Physics",
    month = "5",
    year = "2019"
}

@article{Heinz:2013th,
    author = "Heinz, Ulrich and Snellings, Raimond",
    title = "{Collective flow and viscosity in relativistic heavy-ion collisions}",
    eprint = "",
    archivePrefix = "arXiv",
    primaryClass = "nucl-th",
    doi = "10.1146/annurev-nucl-102212-170540",
    journal = "Ann. Rev. Nucl. Part. Sci.",
    volume = "63",
    pages = "123--151",
    year = "2013"
}

@article{Busza:2018rrf,
    author = "Busza, Wit and Rajagopal, Krishna and van der Schee, Wilke",
    title = "{Heavy Ion Collisions: The Big Picture, and the Big Questions}",
    eprint = "",
    archivePrefix = "arXiv",
    primaryClass = "hep-ph",
    reportNumber = "MIT-CTP-4892, MIT-CTP/4892",
    doi = "10.1146/annurev-nucl-101917-020852",
    journal = "Ann. Rev. Nucl. Part. Sci.",
    volume = "68",
    pages = "339--376",
    year = "2018"
}

@article{ATLAS:2014ipv,
    author = "Aad, Georges and others",
    collaboration = "ATLAS",
    title = "{Measurements of the Nuclear Modification Factor for Jets in Pb+Pb Collisions at $\sqrt{s_{\mathrm{NN}}}=2.76$ TeV with the ATLAS Detector}",
    eprint = "",
    archivePrefix = "arXiv",
    primaryClass = "hep-ex",
    reportNumber = "CERN-PH-EP-2014-172",
    doi = "10.1103/PhysRevLett.114.072302",
    journal = "Phys. Rev. Lett.",
    volume = "114",
    number = "7",
    pages = "072302",
    year = "2015"
}

@article{CMS:2016uxf,
    author = "Khachatryan, Vardan and others",
    collaboration = "CMS",
    title = "{Measurement of inclusive jet cross sections in $pp$ and PbPb collisions at $\sqrt{s_{NN}}=$ 2.76 TeV}",
    eprint = "",
    archivePrefix = "arXiv",
    primaryClass = "nucl-ex",
    reportNumber = "CMS-HIN-13-005, CERN-EP-2016-217",
    doi = "10.1103/PhysRevC.96.015202",
    journal = "Phys. Rev. C",
    volume = "96",
    number = "1",
    pages = "015202",
    year = "2017"
}

@article{ALICE:2013dpt,
    author = "Abelev, B. and others",
    collaboration = "ALICE",
    title = "{Measurement of charged jet suppression in Pb-Pb collisions at $\sqrt{s_{NN}}$ = 2.76 TeV}",
    eprint = "",
    archivePrefix = "arXiv",
    primaryClass = "nucl-ex",
    reportNumber = "CERN-PH-EP-2013-205",
    doi = "10.1007/JHEP03(2014)013",
    journal = "J. High Energy Phys.",
    volume = "03",
    pages = "013",
    year = "2014"
}

@article{ATLAS:2016yzd,
    author = "Aaboud, Morad and others",
    collaboration = "ATLAS",
    title = "{Measurements of long-range azimuthal anisotropies and associated Fourier coefficients for $pp$ collisions at $\sqrt{s}=5.02$ and $13$ TeV and $p$+Pb collisions at $\sqrt{s_{\mathrm{NN}}}=5.02$ TeV with the ATLAS detector}",
    eprint = "",
    archivePrefix = "arXiv",
    primaryClass = "nucl-ex",
    reportNumber = "CERN-EP-2016-200",
    doi = "10.1103/PhysRevC.96.024908",
    journal = "Phys. Rev. C",
    volume = "96",
    number = "2",
    pages = "024908",
    year = "2017"
}

@article{CMS:2015yux,
    author = "Khachatryan, Vardan and others",
    collaboration = "CMS",
    title = "{Evidence for Collective Multiparticle Correlations in p-Pb Collisions}",
    eprint = "",
    archivePrefix = "arXiv",
    primaryClass = "nucl-ex",
    reportNumber = "CMS-HIN-14-006, CERN-PH-EP-2015-011",
    doi = "10.1103/PhysRevLett.115.012301",
    journal = "Phys. Rev. Lett.",
    volume = "115",
    number = "1",
    pages = "012301",
    year = "2015"
}

@article{PHENIX:2017xrm,
    author = "Aidala, C. and others",
    collaboration = "PHENIX",
    title = "{Measurements of Multiparticle Correlations in $d+\mathrm{Au}$ Collisions at 200, 62.4, 39, and 19.6 GeV and $p+\mathrm{Au}$ Collisions at 200 GeV and Implications for Collective Behavior}",
    eprint = "",
    archivePrefix = "arXiv",
    primaryClass = "nucl-ex",
    doi = "10.1103/PhysRevLett.120.062302",
    journal = "Phys. Rev. Lett.",
    volume = "120",
    number = "6",
    pages = "062302",
    year = "2018"
}

@article{Giacalone:2024luz,
    author = "Giacalone, Giuliano and others",
    title = "{Exploiting Ne20 Isotopes for Precision Characterizations of Collectivity in Small Systems}",
    eprint = "",
    archivePrefix = "arXiv",
    primaryClass = "nucl-th",
    reportNumber = "CERN-TH-2024-021",
    doi = "10.1103/k8rb-jgvq",
    journal = "Phys. Rev. Lett.",
    volume = "135",
    number = "1",
    pages = "012302",
    year = "2025"
}

@techreport{Brewer:2021kiv,
    author = "Brewer, Jasmine and Mazeliauskas, Aleksas and van der Schee, Wilke",
    title = "{Opportunities of OO and $p$O collisions at the LHC}",
    eprint = "2103.01939",
    archivePrefix = "arXiv",
    primaryClass = "hep-ph",
    reportNumber = "CERN-TH-2021-028",
    month = "3"
}

@article{Giacalone:2024ixe,
    author = "Giacalone, Giuliano and others",
    title = "{Anisotropic Flow in Fixed-Target $^{208}$Pb+$^{20}$Ne Collisions as a Probe of Quark-Gluon Plasma}",
    eprint = "",
    archivePrefix = "arXiv",
    primaryClass = "nucl-th",
    reportNumber = "CERN-TH-2024-074",
    doi = "10.1103/PhysRevLett.134.082301",
    journal = "Phys. Rev. Lett.",
    volume = "134",
    number = "8",
    pages = "082301",
    year = "2025"
}

@inproceedings{Huang:2023viw,
    author = "Huang, Shengli",
    title = "{Measurements of azimuthal anisotropies in $^{16}$O+$^{16}$O and $\gamma$+Au collisions from STAR}",
    eprint = "2312.12167",
    archivePrefix = "arXiv",
    primaryClass = "nucl-ex",
    month = "12",
    year = "2023"
}

@article{Summerfield:2021oex,
    author = "Summerfield, Nicholas and Lu, Bing-Nan and Plumberg, Christopher and Lee, Dean and Noronha-Hostler, Jacquelyn and Timmins, Anthony",
    title = "{$^{16}$O $^{16}$O collisions at energies available at the BNL Relativistic Heavy Ion Collider and at the CERN Large Hadron Collider comparing $\alpha$ clustering versus substructure}",
    eprint = "",
    archivePrefix = "arXiv",
    primaryClass = "nucl-th",
    doi = "10.1103/PhysRevC.104.L041901",
    journal = "Phys. Rev. C",
    volume = "104",
    number = "4",
    pages = "L041901",
    year = "2021"
}

@article{YuanyuanWang:2024sgp,
    author = "Wang, Yuanyuan and Zhao, Shujun and Cao, Boxing and Xu, Hao-jie and Song, Huichao",
    title = "{Exploring the compactness of {\ensuremath{\alpha}} clusters in O16 nuclei with relativistic O16+O16 collisions}",
    eprint = "",
    archivePrefix = "arXiv",
    primaryClass = "nucl-th",
    doi = "10.1103/PhysRevC.109.L051904",
    journal = "Phys. Rev. C",
    volume = "109",
    number = "5",
    pages = "L051904",
    year = "2024"
}

@article{Choi:2019mip,
    author = "Choi, Suyong and Oh, Hayoung",
    title = "{Improved extrapolation methods of data-driven background estimations in high energy physics}",
    eprint = "",
    archivePrefix = "arXiv",
    primaryClass = "hep-ph",
    doi = "10.1140/epjc/s10052-021-09404-1",
    journal = "Eur. Phys. J. C",
    volume = "81",
    number = "7",
    pages = "643",
    year = "2021"
}

\end{document}